# An Improved Modular Addition Checksum Algorithm

Philip Koopman
Carnegie Mellon University
Pittsburgh, PA, USA
koopman@cmu.edu

*Abstract*—This paper introduces a checksum algorithm that provides a new point in the performance/complexity/effectiveness checksum tradeoff space. It has better fault detection properties than single-sum and dual-sum modular addition checksums. It is also simpler to compute efficiently than a cyclic redundancy check (CRC) due to exploiting commonly available hardware and programming language support for unsigned integer division. The key idea is to compute a single running sum, but introduce a left shift by the size (in bits) of the modulus before performing the modular reduction after each addition step. This approach provides a Hamming Distance of 3 for longer data word lengths than dual-sum approaches such as the Fletcher checksum. Moreover, it provides this capability using a single running sum that is only twice the size of the final computed check value, while providing fault detection capabilities even better than large-block variants of dual-sum approaches that require larger division operations. A variant that includes a parity bit achieves Hamming Distance 4 for the same size check value, approximating the fault detection capabilities of a good CRC for about half the data word length attainable by such a CRC.

I. INTRODUCTION

Modular addition checksums and Cyclic Redundancy Check (CRC) error codes are ubiquitous for guarding the integrity of data against non-malicious corruption, such as faults that occur during data transmission on networks. A perennial tradeoff involves algorithmic complexity, computational cost, and fault detection effectiveness. This work provides a new point in the tradeoff space of better fault detection effectiveness compared to other checksum approaches, with moderate computational cost and algorithmic complexity.

A detailed description of terminology can be found in [Koopman23]. In the interest of brevity the reader is referred to that earlier work for background and terminology details. Earlier work and a more thorough literature review can be found in [Maxino09].

*A. Terminology recap*

A brief recap of terms, shortened from [Koopman23] follows:

- **Code word:** a tuple of data containing a **data word** (the data for which integrity checking is desired) and a **check value** (the result of a checksum calculation). The purpose of using a checksum is to detect whether one or more bit faults in the code word have occurred since it was created. Binary symmetric bit inversion faults (bit flips) can occur in either or both of the data word and the check value.

---

**Modular Single-Sum Checksum:**
  sum = (sum + block[index]) % modulus

**Modular Dual-Sum Checksum:**
  sumA = (sumA + block[index]) % modulus
  sumB = (sumA + sumB) % modulus

**Koopman Checksum:**
  sum = ((sum<<k) + block[index]) % modulus

*Figure 1. Checksum computational kernels. Modulus selection affects fault detection capabilities; k is bits in check value. Koopman checksum moduli are 253, 65519, and 4294967291.*

---

- **Block:** a chunk of data from the data word processed at each step of a summing operation. For a Koopman checksum the block size is the same as the check value size.

- **Modulus:** the unsigned integer used in the checksum calculation kernel to produce a remainder from a division operation. The modulus must be less than or equal to $2^k$ for a k-bit check value to ensure that the result fits within the number of bits used for the check value. Modulus selection can affect checksum effectiveness.

- **Modular checksum:** computes the check value via a summing operation across all the blocks of the data word. Each step in the sum adds a current running sum to the next block value modulo a predefined modulus.

- **Single-sum checksum:** a checksum algorithm that performs a single running modular checksum to compute a check value.

- **Dual-sum checksum:** a checksum algorithm that computes a pair of two running sums in the manner of a Fletcher or Adler checksum and concatenates those sums to produce a check value. (See: [Fletcher82] [Adler].)

- **Large-block checksum:** a checksum algorithm that uses a block size larger than the number of bits in the modulus. For example, a 4-byte block size with a one-byte sized modulus of 255 would be called "d255_b4" to denote it is a dual checksum with modulus 255 and block size 4 bytes.

- **Hamming distance:** the lowest number of faulty bits that has at least one undetectable fault in a code word. HD=2 means all 1-bit faults are detected, but at least one possible 2-bit fault is undetectable. HD=3 means that all 1-bit and 2-bit faults are detected, but there is at least one undetectable 3-bit fault. HD=3 performance for the longest code word possible is desirable for better checksum fault detection effectiveness. Cyclic redundancy checks can often achieve even higher HD values, but are beyond the scope of this paper.
- **Bit Error Ratio (BER):** the fraction of code word bits that are subject to a value inversion corruption (a "bit flip") on a per bit basis. A BER of $10^{-6}$ means that each bit in a code word has an independent one-per-million chance of being flipped. Longer code words have a higher probability of corruption simply because they contain more bits that are each individually subject to the BER.
- **$P_{ud}$:** the probability of an undetected fault for a particular code word given the code word length, the BER, and the checksum algorithm being used. A lower $P_{ud}$ means that fault detection effectiveness is better. All things being equal, longer code words have a higher $P_{ud}$ at the same BER simply because they are more likely to accumulate enough bit faults to meet or exceed the checksum's HD.

*B. Approach*

In this work we introduce a computational wrinkle on a single-sum checksum approach based on insight gained in studying large-block checksums. The result is a modification of a single-sum checksum computational kernel that provides large-block checksum fault detection capabilities via changing the computation into a data-word-long modular reduction operation with a carefully chosen modulus. However, this is done without requiring the use of large integer division arithmetic operations.

The key to the approach is replacing this modular addition core algorithmic step:

   sum = (sum + block) mod m

with a step that promotes better mixing of bits and avoids leakage of single-bit faults from a corrupted block into the sum as a single-bit fault:

   sum = ((sum + block)*$2^k$) mod m

where k is the number of bits required to represent the result of applying modulus m.

The multiplication by $2^k$ is efficiently implemented by a left-shift of k bits, where k is the number of bits in the modulus, typically 8, 16, or 32. From a pseudo-code point of view, the computational kernel is an optimized version of:

   sum = ((sum + block) << k ) % m

where "<<" is a left bit shift and "%" denotes remainder after division.

This formulation is prone to generating carry-out bits that exceed the size of an integer register (e.g., for a 16-bit check value a carry-out bit might result in a 33-bit value). We resolve

---

Initialize $Sum_{initial}$ = InitialSeed

Iterate across each block *i* in data word:
   $Sum_{new}$ = ( $Sum_{old}$ + $Block_i$ ) mod M

Check Value is the final $Sum_{new}$

*Algorithm 1: Single-sum checksum.*

---

this issue by reorganizing the computation to pipeline the modular reduction and addition steps, yielding this computational kernel with an additional first set-up step to prime the pipeline and a final clean-up step to wind down the pipeline:

   sum = ((sum<<k) + block) % m

This formulation is not subject to carry-outs from the addition, and can fit well into typical-size machine words and programming language integer types (e.g., 32-bit integer for a 16-bit check value). A rationale for this approach and analysis are contained in later sections of this paper.

This checksum algorithm provides a new point in the checksum tradeoff space, with algorithmic complexity approaching that of a single-sum checksum and yet having fault detection capability of Hamming Distance (HD) three for useful data word lengths. To be sure, a cyclic redundancy check (CRC) should be used for life critical applications to provide an even higher HD. However, this checksum approach may be attractive where better performance than a regular checksum is desired but the computational cost and algorithmic complexity of implementing an efficient CRC calculation is not warranted.

The remainder of this paper briefly reviews different checksum approaches including this new proposed approach in Section II. Section III explains the conceptual underpinnings of the new approach. Section IV presents fault detection experimental results, and Section V provides conclusions. This paper is intended to be read in conjunction with a previous paper on large-block checksums [Koopman23], and therefore provides only a minimum of context and terminology. The experimental methodology and simulation platform are the same as in that previous work, updated to include this new algorithm.

II. PREVIOUS CHECKSUM ALGORITHMS

Checksum computations of interest for this paper break a data word into blocks, perform modular addition across all the data blocks within the data word to create a check value, and store that check value with the data word to create a code word. That code word can later be checked for integrity by recomputing the checksum from the data word and comparing that result to the stored check value.

Previous checksum algorithms considered include single-sum, dual-sum, and large block size versions of those two algorithms. These are compared against the effectiveness of the newly proposed Koopman checksum algorithm that uses a modified single-sum approach to pipeline a single large modular reduction operation.

## A. Single-Sum Checksums

Classical checksum algorithms involve computing a single modular sum of block values drawn in sequence from the entire length of the data word. A generic description is shown as Algorithm 1.

In Algorithm 1, M is a predetermined algorithm-dependent modulus. Using a k=16 check value (16 bits) as an example, some possible values for the modulus are two's complement addition (65536), one's complement addition (65535), largest prime less than $2^k$ (65521), and a modulus that is suitable for large-block checksums (65525) [Koopman23].

Single sum checksum algorithms can generate any desired k-bit sized check value, with typical values of k being 8, 16, and 32. They provide HD=2 fault detection performance.

## B. Dual-Sum Checksums

A more advanced class of checksums was introduced by Fletcher's work [Fletcher82]. A generic description is shown as Algorithm 2.

---

Initialize $SumA_{initial}$ = $SumB_{initial}$ = InitialSeed

Iterate across each block *i* in data word:
    $SumA_{new}$ = ($SumA_{old}$ + $Block_i$) mod M
    $SumB_{new}$ = ($SumB_{old}$ + $SumA_{new}$) mod M

Check Value is:
    $SumA_{new}$ concatenated with $SumB_{new}$

*Algorithm 2: Dual-sum checksum.*

---

In the Fletcher approach, a pair of running modular sums is used instead of a single sum. The first sum, which we denote SumA, is the same as a single-sum checksum, except sized at only k/2 bits (e.g., an 8-bit running modular sum for a 16-bit check value).

The second sum in Fletcher's algorithm, SumB, is another k/2 bit running sum that is updated not by summing block values, but rather by summing the old version of SumB with the new version of SumA for each block being processed. The check value result is the concatenation of SumA and SumB, each of which are k/2 bits in size, to create a single k-bit check value.

As with Algorithm 1, for Algorithm 2 the modulus is M, with the same modulus being used for both sums. All blocks from the data word are processed in a running sum approach, with the pair of sums updated as each block is processed.

The Fletcher dual-sum checksum algorithm uses a modulus of $2^k$-1 [Fletcher82]. The Adler dual-sum checksum uses a modulus of the largest prime integer less than $2^k$ [Adler]. Large-block checksums can benefit from using an empirically-determined modulus, such as 253 for an 8-bit large-block dual-sum algorithm and 65525 for a 16-bit large-block dual-sum algorithm [Koopman23].

Dual-sum checksums provide HD=3 up to the "rollover" length of the checksum. That rollover length is (((modulus-1)* blocksize)+1) based on the summation process. The longest known 16-bit dual sum checksum rollover point is ((239-1)*14) = 3332 bytes for a 14-byte block size and modulus 239 [Koopman23]. That HD=3 capability comes at the cost of needing to perform a 128-bit integer division operation to generate a division remainder for each 14-byte block of data in the modular reduction process.

## C. Koopman Checksums

This paper introduces the Koopman checksum algorithm shown as Algorithm 3. It is a single-sum modular checksum approach that adds a left-shift by k bits into the running sum process. Since k is typically 8, 16, or 32 bits in practice, this left shift is really just a byte alignment change, making the operation particularly cheap to perform on most computation hardware. (The algorithmic approach should work with any value of k, but k being a multiple of 8 is most likely to be useful in practice.)

---

Initialize $Sum_{initial}$ = $Block_{first}$ ⊕ InitialSeed

Iterate across remaining blocks *i* in data word:
    $Sum_{new}$ = (($Sum_{old}$<<k) + $Block_i$)) mod M

Check Value based on final sum:
    CheckValue = (($Sum_{new}$<<k) mod M)

*Algorithm 3: Koopman checksum. "<<k" is left shift by k bits.*

---

As with Algorithms 1 and 2, M is the modulus. All blocks from the data word are processed in a running sum approach. Since this is a single-sum approach, the value k determines the number of bits in the check value, and the modulus must be less than or equal to $2^k$. In practice, effective moduli choices are 253, 65519, and 4294967291.

This checksum provides HD=3 up to a significant length, longer than that provided by dual-sum checksum approaches. Specifics depend critically on the modulus selected, and will be discussed in a later section.

## D. Checksum Effectiveness Comparison

The experimental framework and $P_{ud}$ calculation approach from [Koopman23] were used. Briefly, specific numbers of bit inversion faults were injected into randomized data word values. The fraction of undetected faults was determined for each m-bit fault (m=1, 2, 3, …) across a sweep of data word lengths spanning 1 to 5120 bytes. That data was used to create a weighted sum of undetected fault probability $P_{ud}$ for each data word length based on the probability of faults with different numbers of bit inversions and the data indicating fraction of undetected m-bit faults.

As can be seen in figure 2, a 16-bit Koopman checksum is dramatically better (lower $P_{ud}$) than either a single-sum checksum or a dual-sum checksum of the same size, detecting all 1-bit and 2-bit faults up to almost 4K byte data words. A 15-bit Koopman checksum combined with a single parity bit provides fault detection capability on a par with a good cyclic redundancy check (CRC) [CRC2023] up to almost 2K bytes, detecting all 1-bit, 2-bit, and 3-bit faults up to that length.

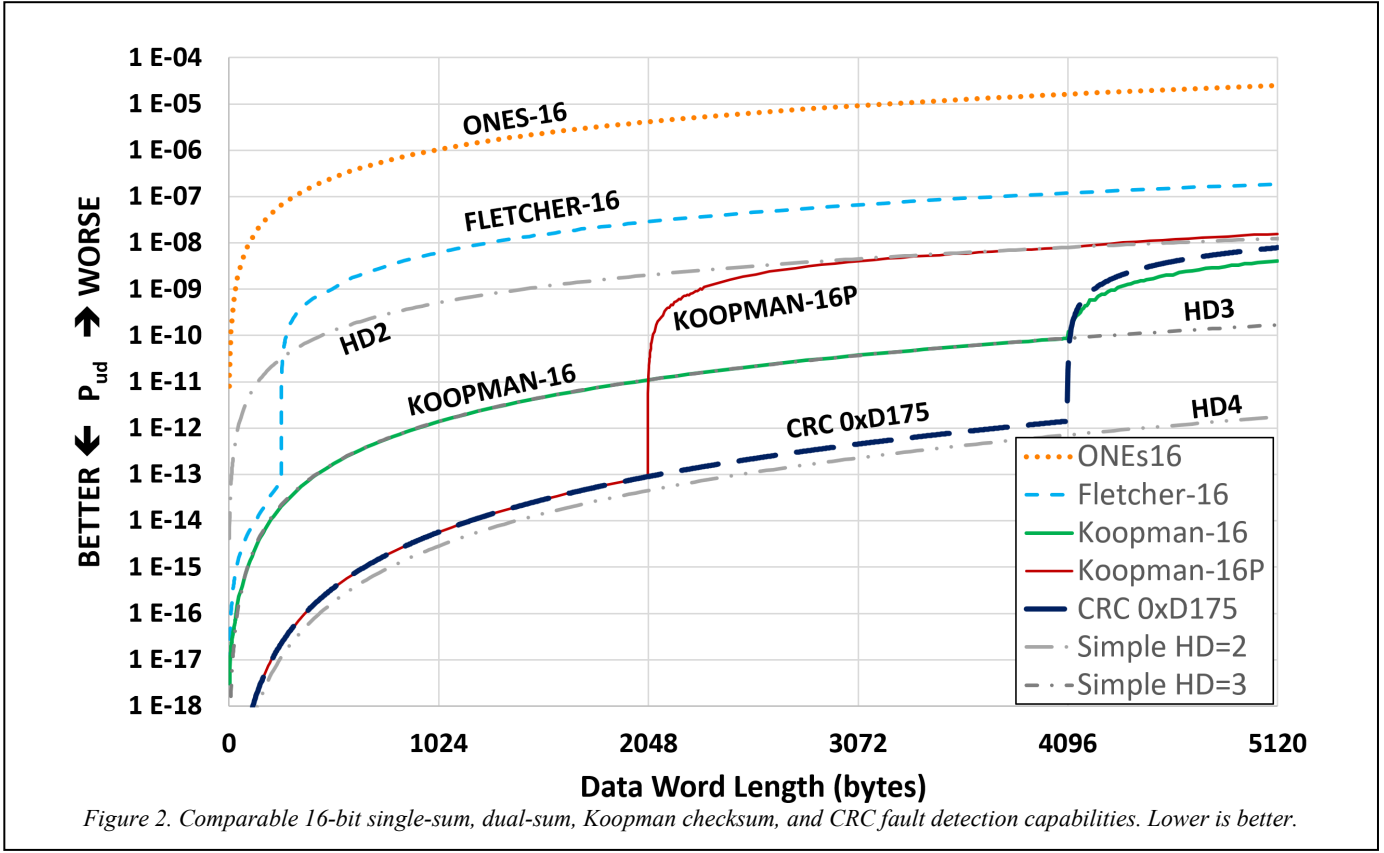

Figure 2. Comparable 16-bit single-sum, dual-sum, Koopman checksum, and CRC fault detection capabilities. Lower is better.

In figure 2, there is no substantial difference between a Fletcher dual-sum checksum with modulus=255 and an Adler dual-sum checksum with modulus=251 at the scale of this plot, so only Fletcher16 is shown. The "Simple HD=4" curve presented as a reference line is a hypothetical idealized checksum with perfect detection of all 1-bit, 2-bit, and 3-bit faults, and $1/2^k$ fraction of undetected faults for all other numbers of bit faults. The Simple HD=2 and HD=3 lines are corresponding idealized curves for Hamming Distances of 2 and 3, respectively.

The following sections of this paper describe the mechanics and theory of operation of the Koopman checksum. They also compare and contrast it to our previous work on large-block checksums. The short version is that a Koopman checksum provides longer HD=3 capability to exploit the full potential of a large-block approach without requiring cumbersome extended-precision integer division operations.

III. KOOPMAN CHECKSUM ALGORITHM ANALYSIS

Decades of increased computer clock speeds and an ever-widening gap between computation speed and memory access delay have fundamentally changed the tradeoffs relevant to computing a checksum value. In particular, the cost of a division operation has been dramatically reduced as a proportion of the time it takes to fetch data from memory, making division comparatively cheaper than it used to be. This is due to a combination of improved native hardware support for division and CPUs that can perform arithmetic faster than memory devices can return large amounts of data – cache memory hierarchies notwithstanding. Additionally, it is more common to see more capable CPUs used in embedded applications that provide efficient support for 32-bit or even 64-bit division operations.

These trends argue that there is a place in the checksum speed/effectiveness tradeoff space for an algorithm that uses a division operation more aggressively than traditional modular checksum approaches, while being more streamlined than large-block approaches recently discovered. The result is something that provides better bit mixing than addition-centric approaches, does not require using huge integer division operations, and yet avoids the full complexity of a Cyclic Redundancy Check (CRC) computation. Notably, the proposed algorithm makes use of hardware support for integer division operations to perform bit mixing rather than requiring software-only support for CRC-based bit mixing.

An explanation of the algorithm is presented in two parts: (1) using an offset to mitigate 2-bit fault vulnerabilities for large-block checksums, and (2) an iterative approach to performing modular reduction operation on an unbounded length data word.

A. *Mitigating large-block 2-bit faults*

As shown in our previous work, large-block checksums can significantly improve the fraction of undetected faults, but are limited to HD=2 in practice unless combined with another technique such as a dual-sum approach [Koopman23].

Further exploration has revealed that the HD=2 behavior for large-block single-sum checksums was largely attributable to poor mixing of the lowest few bits of the block by the modulus. We illustrate this with a concrete example using an 8-bit modulus and a 32-bit block in Table 1. ("0x" denotes a hexadecimal value.)

It is useful to consider the implications of Table 1 for a data word that is all zero, with the one set bit in the values in that table representing how a single bit flip injected fault affects the modular reduction result. (The principles apply more generally to other values in practice.)

| Block Value | Mod M | Result | |
|---|---|---|---|
| 0x80000000 | % 0xFD | 0xA7 | |
| 0x40000000 | % 0xFD | 0xD2 | |
| 0x20000000 | % 0xFD | 0x69 | |
| 0x10000000 | % 0xFD | 0xB3 | |
| 0x08000000 | % 0xFD | 0xD8 | |
| 0x04000000 | % 0xFD | 0x6C | |
| 0x02000000 | % 0xFD | 0x36 | |
| 0x00100000 | % 0xFD | 0x1B | |
| 0x00800000 | % 0xFD | 0x8C | |
| 0x00400000 | % 0xFD | 0x46 | |
| 0x00200000 | % 0xFD | 0x23 | |
| 0x00100000 | % 0xFD | 0x90 | |
| 0x00080000 | % 0xFD | 0x48 | |
| 0x00040000 | % 0xFD | 0x24 | |
| 0x00020000 | % 0xFD | 0x12 | |
| 0x00010000 | % 0xFD | 0x09 | |
| 0x00008000 | % 0xFD | 0x83 | |
| 0x00004000 | % 0xFD | 0xC0 | |
| 0x00002000 | % 0xFD | 0x60 | |
| 0x00001000 | % 0xFD | 0x30 | |
| 0x00000800 | % 0xFD | 0x18 | |
| 0x00000400 | % 0xFD | 0x0C | |
| 0x00000200 | % 0xFD | 0x06 | |
| 0x00000100 | % 0xFD | 0x03 | |
| *0x00000080* | *% 0xFD* | *0x80* | ** |
| *0x00000040* | *% 0xFD* | *0x40* | ** |
| *0x00000020* | *% 0xFD* | *0x20* | ** |
| *0x00000010* | *% 0xFD* | *0x10* | ** |
| *0x00000008* | *% 0xFD* | *0x08* | ** |
| *0x00000004* | *% 0xFD* | *0x04* | ** |
| *0x00000002* | *% 0xFD* | *0x02* | ** |
| *0x00000001* | *% 0xFD* | *0x01* | ** |

Table 1. Example 4-byte block with one-bit values modulo 253. Block values with a single non-zero bits in the lowest byte are unchanged. Rows marked with "**" are unchanged by a modular reduction operation.

In Table 1, a one-bit fault in block positions larger than the modulus (bit positions 8 through 31) is mixed sufficiently well that no single-bit result is generated. This means that at least three bits will need to be flipped to create an undetectable fault if there is only one block in the data word: one bit flip in the data block and two corresponding bits in the check value. No two-bit fault can go undetected, again assuming there is only this one block in the data word.

However, bits in the block that are in positions smaller than the modulus (bit positions 0 through 7) are not mixed at all, and go through the modular reduction unchanged. This creates a significant HD=2 vulnerability mechanism for single-sum checksums.

While this pass-through behavior for low order bit positions is exactly as expected per the mathematical properties of modular reduction, it makes any operation of this type vulnerable to two-bit faults, albeit with reduced probability compared to a small-block approach. The vulnerability is reduced proportionally with larger block sizes, accounting for somewhat better fault detection effectiveness for larger block sizes seen in figure 2 for Add65525_b4 (single-sum addition with modulus of 65525 and block size of 4 bytes).

A similar effect occurs for large-block dual-sum checksums, but is complemented by the HD=3 effectiveness of the dual-sum approach. For dual-sum checksums, the use of larger blocks extends the HD=3 distance as shown by the difference between the Fletcher checksum (modulus of 255, block size of 1 byte) and the Dual253_b4 checksum (modulus of 253, dual-sum, block size of 4 bytes).

The more important observation is that we are actually getting HD=3 for all the bytes without needing a dual-sum approach – except that pesky low order byte. All the results for single-bit values in higher order bit positions result in two or more bits in the modular reduction result, ensuring HD=3 or better so long as the checksum operation only involves a single block.

We can remove this HD=2 vulnerability by avoiding the use of the lowest eight bits of the block, and forcing them to zero regardless of the data word. This means that what was formerly a 4-byte block in the stated example now holds 3 bytes from the data word and a low byte value that has been forced to be zero. In other words, we are changing to a 3-byte block from the point of view of the checksum, while using a 4-byte modular addition computation.

Making this concrete, for a byte-organized data word starting at byte i, a single step in the checksum calculation used to be for a 4-byte block:

block = (dw$_i$ * 0x1000000) + (dw$_{i+1}$ * 0x10000)
          + (dw$_{i+2}$ * 0x100) + (dw$_{i+3}$ * 0x1)
sum = (sum + block) % modulus

In this notation dw$_i$ is the *i*th byte of the data word, with *i* incremented by the block size of 4 for each iteration through the summation kernel.

To avoid the vulnerability to bit faults affecting the lowest byte, this changes to a 3-byte block calculation:

block = (dw$_i$ * 0x1000000) + (dw$_{i+1}$ * 0x10000)
          + (dw$_{i+2}$ * 0x100)
sum = (sum + block) % modulus

where the byte dw$_{i+3}$ that used to be included in this block is now moved into the next block instead, and the bottom eight bits are left as a value of zero. The index is incremented by 3 for each block processed, resulting in an effective block size of 3. No bytes are skipped, but more iterations of the checksum are required to process all blocks.

Making this change results in an HD=3 checksum algorithm up to the limit of the modulus' capability to support large blocks. In the case of modulus 253, that limit is 13 bytes. To be sure, we are throwing away one byte of capability by not using one byte in the block, so this approach scales to 12-byte blocks from the data word, using a 13-byte modular addition operation having a zero value for the lowest byte. A modulus of 239 has the maximum rollover limit of a 13-byte block using a 14-byte block, but is further away from being able to use the full range of 8 bit check values.

*B. Unbounded length modulus computation*

While setting the lowest byte to zero in the checksum kernel is a nice idea, it doesn't really help for longer data words that require multiple blocks. It is possible in general for two single-bit faults, each in a different block, to induce multi-bit changes to the sum that end up canceling each other out when added within a checksum computation.

To reap the full benefit of zeroing the bottom block byte we need to ensure that the checksum operation uses only a single block. The block size must be the same as the data word size, even for large data words. For 8-bit checksums this is less of an issue, since 13-byte blocks is the upper limit of HD=3 performance, and that can fit in an admittedly large 128-bit integer. However, managing integers larger than that becomes inconvenient in most software environments, making this approach as it is not particularly useful for 16-bit or 32-bit check values that might guard the integrity of data words having thousands or millions of bytes of data.

This issue can be resolved by restructuring the computation to be scalable to an unbounded data word sizes without needing software support for large-precision division operations. We do this by breaking the modular reduction operation down into an iterated process.

Consider the example 32-bit integer value 0x12345600. This value has been chosen so that each 4-bit value and its position is distinctive without loss of generality. The bottom 8 bits have been set to zero in keeping with the strategy of avoiding pass-through of single bit faults, so this represents three bytes of the data word in a block with value "123456" and a byte of zeros at the bottom to ensure HD=3 fault detection effectiveness when undergoing modular reduction.

The progression below is simply a reorganization of the modular reduction operation to reduce it in size from a single 32-bit division to a set of three 16-bit divisions, exploiting the modulus operation property that (A+B) mod M = ((A mod M) + (B mod M)) mod M. Note that "0x" is omitted after the first line below for cleaner notation, but all values in the example are hexadecimal values.

```
0x12345600 mod 0xFD = 0xC8

= (   (12000000 mod FD)
    + (00340000 mod FD)
    + (00005600 mod FD)
    + (00000000 mod FD)) mod FD

=     ((12 * 1000000) mod FD
    +  (34 * 10000) mod FD
    +    (56 * 100) mod FD
    +          (00) mod FD) mod FD

= ( (((12 * 100) * 100) * 100) mod FD
    +      ((34 * 100) * 100) mod FD
    +             (56 * 100) mod FD
    +                   (00) mod FD) mod FD
```

This can be rearranged to:
```
= (((((((12
        * 100) + 34) mod FD)
        * 100) + 56) mod FD)
        * 100) + 00) mod FD
```

Rewriting this into the style of source code, the computation becomes:

```
sum = 0x12;
sum = ((sum * 0x100) + 0x34) % 0xFD;
sum = ((sum * 0x100) + 0x56) % 0xFD;
sum = ((sum * 0x100) + 0x00) % 0xFD;
```

This is a specific example of a computation expressed more generally as a 16-bit implementation that generates an 8-bit check value shown in Appendix A. The first and last line of code are performed only once, but the middle lines are iterated as many times as necessary to process all the block in the data word. The example is arranged so that the reader can easily see how the original block value of 0x12345600 has been distributed across the computation into a byte-by-byte modular addition operation.

Given that multiplication by 0x100 is the same as an 8 bit left shift, from a software implementation point of view, this reformulation amounts to repeated applications of a shift left by 8 bits and add in the next byte, sketched out below:

```
sum = 0x12;
sum = ((sum << 8) + 0x34) % 0xFD;
sum = ((sum << 8) + 0x56) % 0xFD;
sum = ((sum << 8)) % 0xFD;
```

The last byte being processed is always zero to avoid vulnerability to a one-bit fault evading the modular reduction process, so the last step omits the addition of zero, underscoring that it is a clean-up step to wrap up the pipelined implementation.

This example never requires more than 16 bits to hold an intermediate value (twice the check value size of 8 bits).

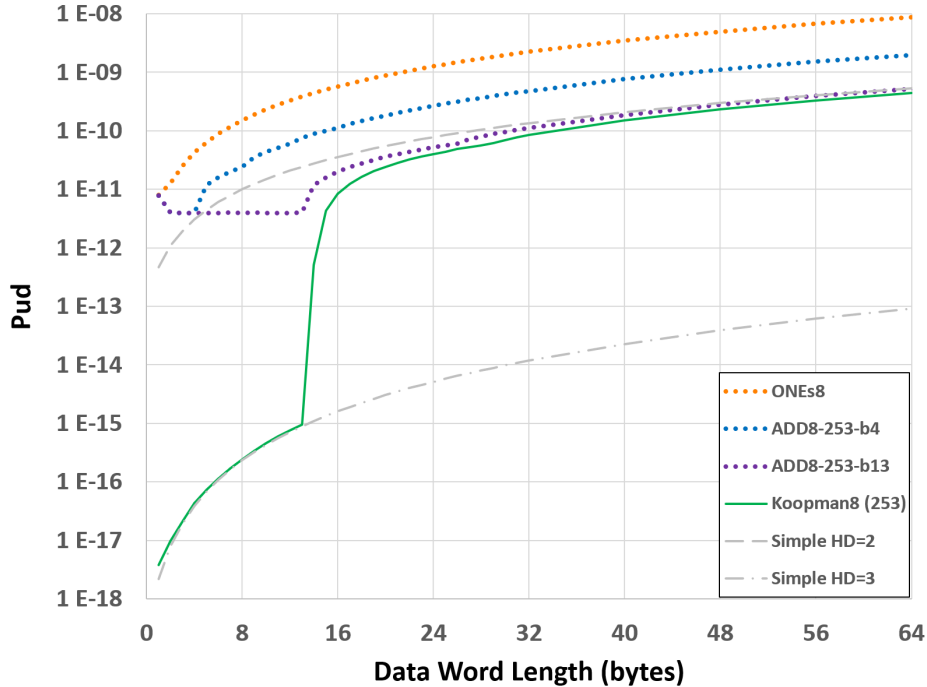

*Figure 3. 8-bit checksum performance comparison including large-block algorithms.*

Moreover, the only division required is an unsigned 16-bit divided by 8 bit division resulting in an 8-bit remainder value.

From a carry-out and overflow point of view, the modulo operation forces each sum to fit into an 8 bit range. That ensures there will be no overflow when the sum from a previous step is shifted left. Additionally, since each block is limited to 8 bits, there can be no overflow from the 16 bit value resulting from the addition, nor any carry-out from the low 8 bits into the high 8 bits of the addition operation.

The computational cost for generating a k-bit check value is one addition and one division operation per k-bit processed element of the data word. Both operations require a k*2 bit holding register for the sum. In some systems a 2*k bit dividend with a k bit divisor such as the one required can be computed more cheaply than a full 2*k-bit division.

More generally, the processing block size can be independent of the check value size [Koopman24]. The left shift in each iteration needs to match the block size, but that need not match the check value size. For example, a 7-bit check value can be processed 8 bits at a time with an 8-bit shift in each iteration. The holding register size in this case must be k bits plus the size of one block so as to fit both the modulo result and the next block to be processed without overlapped bits. The main constraint is that at least k bits of zeros must be included at the end of the computation even if the block size is less than k.

## IV. FAULT DETECTION EFFECTIVENESS

### A. Rollover limit

Any Koopman checksum has a rollover length that limits its HD=3 effectiveness.

There are two mechanisms that affect the rollover length. The first is a pair of bit flips in the data word that just happen to "cancel" each other out, resulting in no change to the modulo result. The second mechanism is a single bit flip in the data word that just happens to result in a single bit changed in the check value, which might also be subject to a second bit flip, resulting in a corrupted check value matching the modulo result of the corrupted data word. Every modulus will eventually suffer one of these two issues with increasing data word length. The question is, which moduli to choose to provide the longest HD=3 capability.

Note that this is a different rollover mechanism than the one present in dual-sum checksums, which is based overflows of accumulated values from being represented by the sumB value in that computation [Fletcher82].

Modulus evaluation was done by using a screening tool that tried all possible combinations of two-bit faults at all lengths up to the reported HD=3 length. For moduli of 16 bits and smaller, all possible moduli were screened. For larger moduli the highest several hundred moduli were screened, so it is possible that a better modulus exists, but the moduli reported are very likely to be among the best, and are possibly the best.

### B. Koopman8 checksum effectiveness

Figure 3 shows the fault detection effectiveness of an 8-bit check value implementation. The best 8-bit modulus is 239, with a rollover length of 14 bytes. This provides HD=3 performance up to 13 byte data words. However, the modulus of 253 is also viable. It is slightly less appealing due to its rollover length of 13 bytes. However, it has a benefit of more effective use of the check value by supporting a range of values {0..252} rather than the smaller range of {0..238}. We recommend 253 unless it is crucial to have that one byte longer HD=3 data word length afforded by modulus 239. These moduli recommendations are the result of brute-force screening campaign of all moduli 128 through 255.

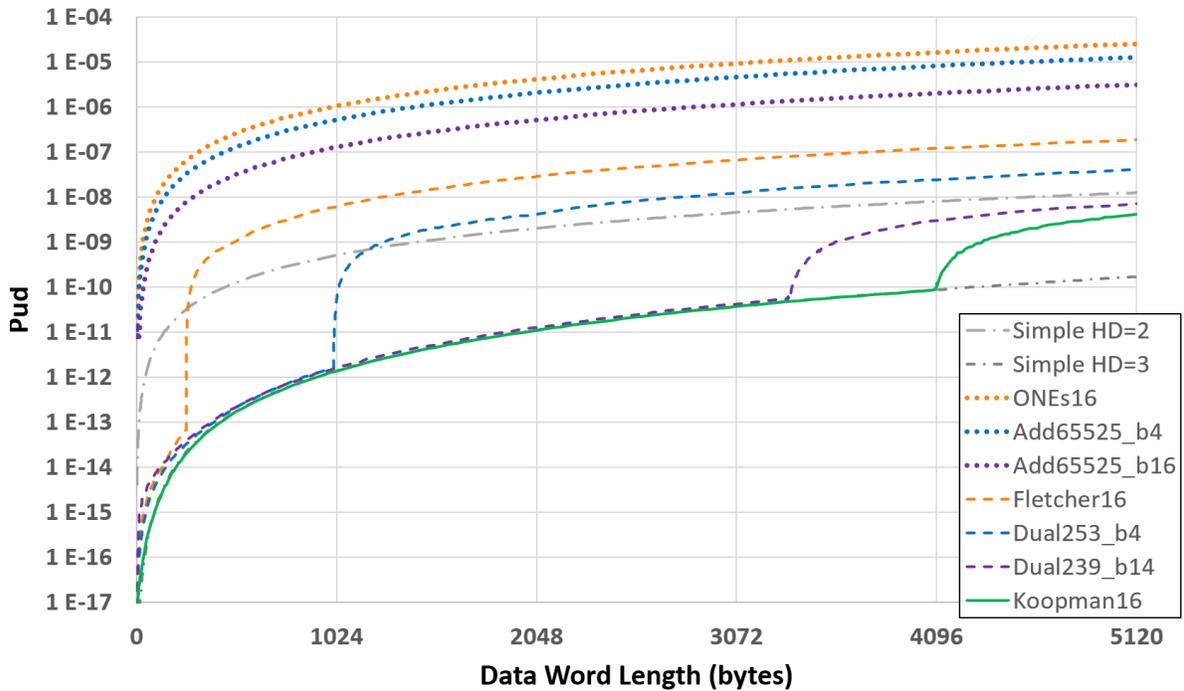

*Figure 4. 16-bit checksum performance comparison including large-block algorithms.*

Data was not available for plotting an 8-bit Fletcher checksum due to the dual 4-bit sums not being compatible with our byte-oriented simulation tooling. However, a Fletcher checksum with a modulus of 15 would have a rollover length of fifteen 4-bit nybbles, providing 7 bytes of HD=3 capability. That means a Koopman8 checksum provides HD=3 performance for longer data word lengths than an 8-bit Fletcher checksum.

Comparing the 8-bit Koopman checksum to a large-block 8-bit single-sum checksum reveals that for long data word lengths the $P_{ud}$ is close to a very large block checksum – which makes sense. Both are achieving HD=2 for large data word lengths with the Koopman checksum effectively being the maximum possible block length, equal to the data word length. Thus, even at long data word lengths, a Koopman checksum provides the benefits of a large-block checksum without the need to perform large integer division operations.

### C. Koopman16 checksum effectiveness

The 16-bit version of the Koopman checksum uses a modulus of 65519. This provides HD=3 up to data words of 4092 bytes. At 4094 and higher data word lengths it provides HD=2. This modulus was selected as providing the longest HD=3 length based on an exhaustive search of all 16-bit moduli.

Figure 4 shows simulation results for fault detection effectiveness. Koopman16 checksum performance is better than even the maximum-length large block dual sum checksum (modulus of 239, block size of 14), even though it only needs a 16-bit division instead of the 128-bit division required in practice by the dual checksum algorithm. The much larger $P_{ud}$ gap between the Dual253_b4 and Koopman16 Pud curves is more of a fair comparison in terms of implementation, because both use 32-bit division operations.

### D. Koopman32 checksum effectiveness

The recommended 32-bit modulus is 4294967291 (which is 0xFFFFFFFB, and thus close to $2^k$). This has a maximum HD=3 length of 134,217,720 bytes.

While simulation resources do not permit comprehensive experimental confirmation of the 32-bit modulus, some limited experiments failed to find any undetected 2-bit faults, including 50 million trials at a 1 MB data word length, and one million trials at the cancellation length. This suggests that the modulus is worth provisional use pending further investigation.

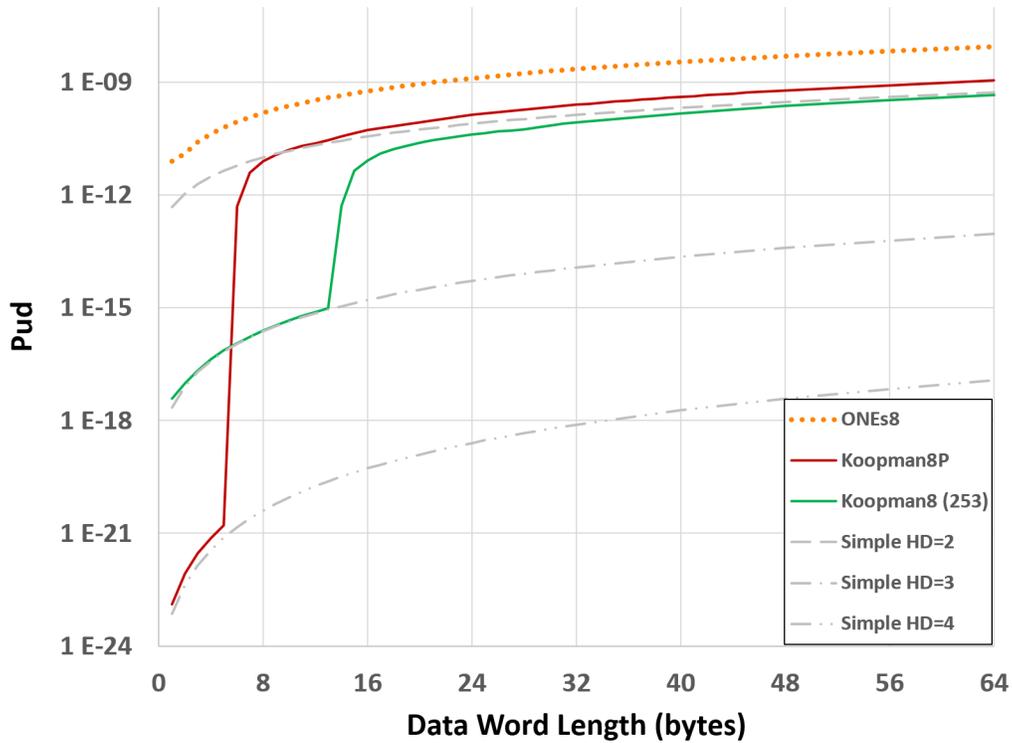
*Figure 5. Comparison of Koopman8 and Koopman8P checksums.*

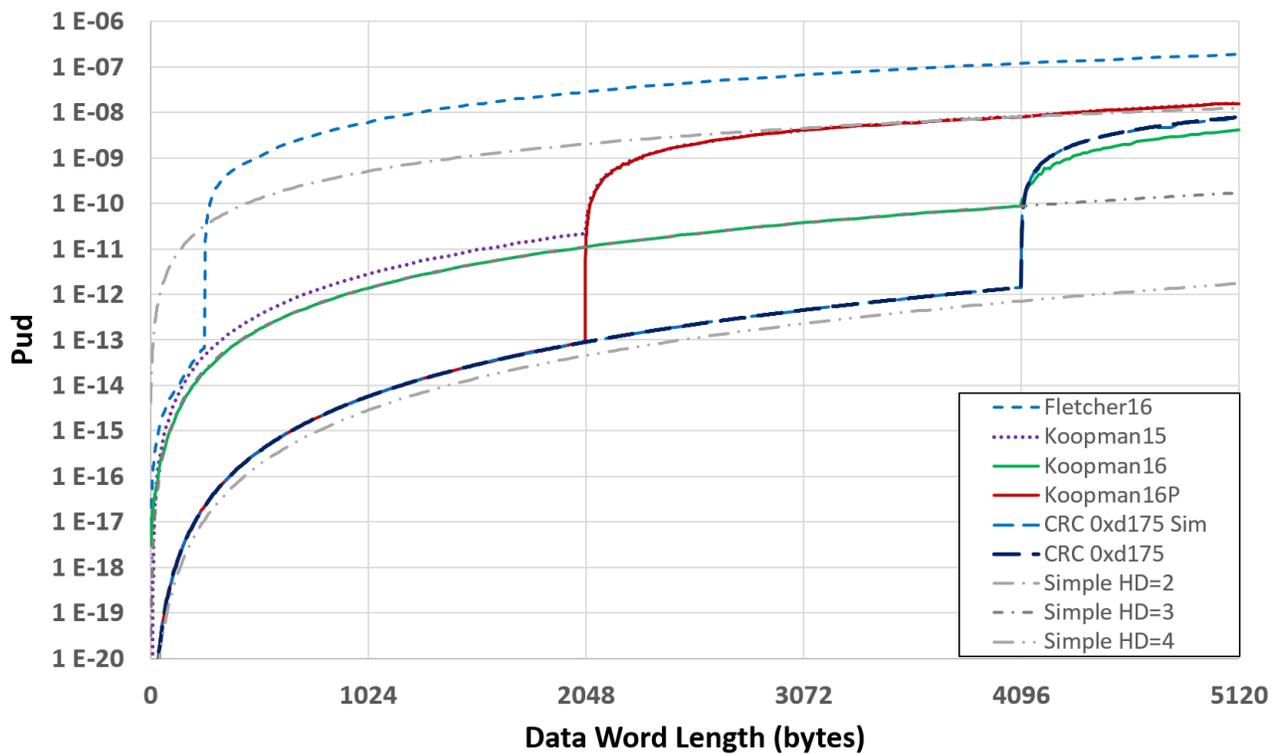
*Figure 6. Comparison of Koopman15, Koopman16 and Koopman16P checksums.*

## V. CHECKSUM PLUS PARITY VARIANT

A variant of the Koopman checksum concatenates a k-1 bit checksum with a parity bit to achieve an HD=4 k-bit check value, albeit at shorter lengths than an HD=3 checksum for a given check value size.

The data word is still processed in k-bit blocks. The checksum kernel is doing in effect one very large modulo operation processed k bits at a time, so a smaller-than-k-bit modulus does not change this aspect of the algorithm. A k-1 bit modulus is used to leave one bit unused in the check value for a parity bit.

Using the same selection criteria as for a k-bit modulus, good 7, 15, and 31 bit moduli are as follows:

- Modulus 125: Provides HD=4 up to 5 byte data word.
- Modulus 32749: Provides HD=4 up to 2044 byte data word
- Modulus 2147483629: Provides HD=4 up to 134217720 bytes.

The reason this works is that parity detects all odd number of bit faults. Using an HD=3 k-1 bit Koopman checksum modulus (e.g., 15-bits), produces a check value that fits into k-1 bits. The remaining $k^{th}$ can then hold a parity bit, bumping the composite HD=3 up to HD=4 by having the parity bit catch all the three-bit faults missed by the Koopman checksum.

Using k=16 bits as an example, the Koopman16P HD=4 performance is only about half as long as Koopman16 checksum HD=3 performance, because a 15-bit modulus rolls over at about half the length of a 16-bit modulus, giving HD=2 above that rollover length. However, HD=4 gives dramatically better fault detection $P_{ud}$ for a BER-based fault model, so having an HD=4 checksum can be useful for some applications that only need integrity checks for that somewhat shorter data word length.

The implementation shown in Appendix A.c XORs the block bits together to compute parity of each bit position for speed, and defers boiling the result down to a single parity bit until the end. The final one-bit shift left of the sum value is to make room to place the parity bit in the bottom bit of the check value. (The placement of the parity bit might instead be at the top of the check value without changing fault detection effectiveness.) An important detail is that the final check value from the checksum calculation proper must be included in the parity computation to catch any bit flips that occur to the check value itself. This provides a complete HD=4 integrity check at appropriately-sized data word lengths.

Figure 5 shows the comparative performance of Koopman8 (8-bit) vs. Koopman8P (7-bit Koopman checksum with a parity bit placed in the $8^{th}$ bit of the check value) checksums. Figure 6 shows the comparative performance of Koopman16 and Koopman16P checksums. In both cases the Koopman+parity checksum achieves HD=4 for approximately half the HD=3 length of the non-parity checksum with the same check value size.

## VI. CONCLUSIONS

This paper makes the following contributions:

1. A novel single-sum modular checksum approach with unbounded block size provides fault detection effectiveness of HD=3 for useful data word lengths. A key idea is inserting zeros in the bottom bits of the block before the modulo operation so that all non-zero bits are in bit positions higher than the modulus value.

2. The HD=3 capabilities of Koopman checksums are longer than dual-sum checksums such as the Fletcher checksum, even when a large-block approach is used. This is accomplished while using smaller sized integers that are only twice the check value size.

3. An efficient iterated algorithm computes a modulo operation up to an unbounded data word length. The computational kernel requires an iterated 2*k bit unsigned division producing a k-bit remainder for each k-bit block processed (typically k=8, 16, or 32).

4. The following moduli are identified as good candidates for use with this checksum approach: 253, 65519, and 4294967291 for Koopman8, Koopman16, and Koopman32 checksums, respectively. Modulus 239 is a useful alternative in special cases that require its HD=3 capability of 13 bytes.

5. A hybrid strategy, using a k-1 bit Koopman checksum and an additional parity bit, is identified that provides HD=4 fault detection effectiveness for approximately half the length of a comparable k-bit checksum HD=3 fault detection capability. Good candidate moduli are: 125, 32749 and 2147483629 for Koopman8P, Koopman16P, and Koopman32P checksums, respectively.

It is important to remember that Cyclic Redundancy Checks (CRCs) can provide superior fault detection mechanisms to even these improved checksum approaches. Nonetheless, for situations in which CRCs are too complex or too slow, this new checksum approach can provide better fault detection effectiveness than previously known modular checksum approaches.

A more extensive treatment of the Koopman Checksum algorithm can be found in Chapters 7, 8 and 11 of [Koopman24].

## VII. REFERENCES

[Adler] Wikipedia, Adler-32, http://en.wikipedia.org/wiki/Adler-32, Dec. 2005.

[C++] "<cstdint> (stdint.h)," https://cplusplus.com/reference/cstdint/ accessed Feb 25, 2023.

[CRC2023] Koopman, P., Best CRC Polyonmials, https://users.ece.cmu.edu/~koopman/crc/ accessed 5/31/2023.

[Fletcher82] Fletcher, J. G. (January 1982). "An Arithmetic Checksum for Serial Transmissions". IEEE Transactions on Communications. COM-30 (1): 247–252. https://doi.org/10.1109/TCOM.1982.1095369

[Koopman23] Koopman, P. "Large-Block Modular Addition Checksum Algorithms," preprint, 2023. https://arxiv.org/abs/2302.13432

[Koopman24] Koopman, P., Understanding Checksums and Cyclic Redundancy Checks, ISBN-13: 979-8876380579, 2024.

[Maxino09] Theresa C. Maxino, Philip J. Koopman (January 2009). "The Effectiveness of Checksums for Embedded Control Networks," IEEE Transactions on Dependable and Secure Computing. https://doi.org/10.1109/TDSC.2007.70216

*Editorial notes:*
- *This version of the paper recommends different moduli than some previous versions due to the availability of a more precise evaluation approach. The moduli recommended in previous versions remain effective as advertised. These new moduli are better.*
- *The naming convention has been updated: Koopman16P refers to a 16-bit check value with a 15-bit modulus plus one parity bit.*
- *The use of a non-zero initial seed has been added to algorithm descriptions.*

APPENDIX A: EXAMPLE CODE

The below C code fragments are intended to illustrate the key idea behind the use of Koopman checksums. They are written in a way to make the key ideas obvious. They are not intended as an illustration of portability or otherwise-desirable code structure.

Variable typing is per <cstdint>, <stdint.h>, or a similar definition approach [C++]. The variable dwSize is assumed to be the number of relevant elements in an 8-bit data word array, although any size block can be processed so long as the intermediate sum variable is large enough to hold the next block plus the current modulo result without overlap.

*A. Koopman8 checksum*

```
uint8_t Koopman8(uint8_t dataWord[],
        uint32_t dwSize, uint32_t modulus)
{
  assert((modulus == 253) || (modulus == 239));
  assert(dwSize > 0);
  assert(initialSeed <= 0xFF);

  uint32_t sum = dataWord[0] ^ initialSeed;

  for(uint32_t index = 1; index < dwSize; index++)
  {
    sum = ((sum<<8) | dataWord[index]) % modulus;
  }

  // Append implicit zero
  sum = (sum<<8) % modulus;
  return((uint8_t)sum);
}
```

*B. Koopman16 checksum; 8-bit blocks*

```
uint16_t Koopman16(uint8_t dataWord[],
        uint32_t dwSize, uint32_t modulus)
{
  assert(modulus == 65519);
  assert(dwSize > 0);
  assert(initialSeed <= 0xFF);

  uint32_t sum = initialSeed ^ dataWord[0];

  for(uint32_t index = 1; index < dwSize; index++)
  {
    sum = ((sum<<8) + dataWord[index]) % modulus;
  }

  // Append two bytes of implicit zeros
  sum = (sum<<8) % modulus;
  sum = (sum<<8) % modulus;
  return((uint16_t)sum);
}
```

*C. Koopman16P checksum*

This implementation assumes there is a function "Parity()" which returns a 1-bit parity value of the input.

```
uint16_t Koopman16P(uint8_t dataWord[],
        uint32_t dwSize, uint32_t modulus)
{
  assert(modulus == 32749);
  assert(dwSize > 0);
  assert(initialSeed <= 0xFF);

  uint32_t sum = initialSeed ^ dataWord[0];
  uint32_t psum = sum; // Initialize parity sum

  for(uint32_t index = 1; index < dwSize; index++)
  {
    sum = ((sum<<8) + dataWord[index] ) % modulus;
    psum ^= dataWord[index];
  }

  // Append two bytes of implicit zeros
  sum = (sum<<16) % modulus;

  // Pack sum with parity
  sum = (sum<<1) | Parity((uint8_t)psum);

  // Append parity as bottom bit of check value
  return((uint16_t)sum);
}
```

*D. Koopman32 checksum*

```
uint32_t Koopman32(uint8_t dataWord[],
        uint32_t dwSize, uint32_t modulus)
{
  assert(dwSize > 1);
  assert(modulus == 4294967291);
  assert(initialSeed <= 0xFF);

  uint64_t sum = initialSeed ^ dataWord[0];

  for(uint32_t index = 1; index < dwSize; index++)
  {
    sum = ((sum<<8) + dataWord[index]) % modulus;
  }

  // Append four bytes of implicit zeros
  sum = (sum<<32) % modulus;
  return((uint32_t)sum);
}
```

*E. Koopman32P checksum*

    This implementation assumes there is a function "Parity()" which returns a 1-bit parity value of the input.

```
uint32_t Koopman32P(uint8_t dataWord[],
          uint32_t dwSize, uint32_t modulus)
{
  assert(dwSize > 1);
  assert(modulus == 0x7FFFFFED);

  uint64_t sum = initialSeed ^ dataWord[0];
  uint32_t psum = (uint32_t)sum; // Initialize parity sum

  for(uint32_t index = 1; index < dwSize; index++)
  {
    sum = ((sum<<8) + (uint64_t)dataWord[index] )
                      % modulus;
    psum ^= dataWord[index];
  }

  // Append four bytes of implicit zeros
  sum = (sum<<32) % modulus;

  // Pack sum with parity
  sum = (sum<<1) | Parity((uint8_t)psum);

  // Append parity as bottom bit of check value
  return((uint32_t)sum);
}
```